\documentclass[prb,twocolumn]{revtex4-1}
\usepackage[T1]{fontenc}
\usepackage{amsmath}
\usepackage{amssymb}
\usepackage{epsfig}

\newcommand{\Tio}{TiO\ensuremath{_2}}

\newcommand{\zero}{\ensuremath{\langle 001 \rangle}}
\newcommand{\zeroa}{\ensuremath{\langle 001 \rangle_{O}}}
\newcommand{\zerob}{\ensuremath{\langle 001 \rangle_{Ti}}}
\newcommand{\oneone}{\ensuremath{\langle 1\bar{1}1 \rangle}}
\newcommand{\oneoner}{\ensuremath{\langle 1\bar{1}1 \rangle_{R}}}

\begin{document}

\title{Steps on Rutile \Tio(110): Active Sites for Water and Methanol Dissociation }
\author{Umberto Martinez, Lasse B.\ Vilhelmsen,  Henrik H.\ Kristoffersen,  Jess Stausholm-M{\o}ller and Bj{\o}rk Hammer}
\affiliation{Interdisciplinary Nanoscience Center (iNANO) and Department of Physics and Astronomy, Aarhus University, DK-8000 Aarhus C, Denmark}
\date{\today}

\begin{abstract}
We present a detailed investigation of the structure and activity of extended defects namely monoatomic steps on  (1$\times$1)--TiO$_2$(110). 
Specifically, the two most stable $\langle 001 \rangle$ and $\langle 1\bar{1}1 \rangle$ step edges are considered. 
Employing an automated genetic algorithm that samples a large number of candidates for each step edge, more stable, reconstructed structures were found for the $\langle 1\bar{1}1 \rangle$ step edge, while the bulk truncated structures were recovered for the $\langle 001 \rangle$\ step edge. 
We demonstrate how oxygen vacancies along these defects have lower formation energies than on flat terraces and how water and methanol molecules adsorb dissociatively on reduced \oneone\ step edges.
Our findings are in agreement with earlier experimental results and indicate an important contribution from step edges to the reactivity of the TiO$_2$(110) surface.
\end{abstract}

\maketitle

\section{Introduction}

Titanium dioxide has attracted much attention due to its catalytic and photocatalytic properties, which allow a wide range of applications \cite{Valden1998,Fujishima2008,Henderson2011}.
Properties which can be extremely different depending on the morphology that may range from nanoparticles of various shape and size to porous films to polycrystalline materials.
All these systems have a large fraction of atoms at steps, edges and corners, which in the case of metal and semiconductors are believed to be active sites  \cite{Zambelli1996,Jeong1999,Hendriksen2010}.
Although such extended defects play an important role also in the physical and chemical properties of metal oxide surfaces, much less attention has been devoted to these systems \cite{Gong2006}.
For example, step edges on metal oxides can act as nucleation sites for deposited metal clusters \cite{Chen2000} and are important in growth mechanisms and surface reconstructions \cite{McCarty2003,McCarty2003a}.

Moreover, it is of fundamental importance to understand the detailed mechanism of the interaction between \Tio\ surfaces with molecules that are present under real conditions (humid air, liquid environment, etc.) \cite{Aschauer2011}.
Water is one of the most studied molecules interacting with \Tio\ surfaces owning to its photocatalytic applications \cite{Diebold_report,Henderson2011}.
Even in surface science studies that use UHV pressures water molecules are always present.
On the other hand, alcohols interacting with \Tio\ surfaces are model systems for the catalytic and photocatalytic  oxidation of organic contaminants \cite{Diebold_report} and methanol, being the simplest one, has  often been used as probe molecule \cite{Henderson1999,Zhou2011}.
 
Here we study the rutile \Tio(110) surface which is considered an important model system for metal oxide surfaces  \cite{Diebold_report,Dohnalek2010,Henderson2011}.
In particular the unreconstructed (1$\times$1) surface is the  thermodynamically most stable face of rutile and is expected to be present on every sample, from nanoparticles to  polycrystalline aggregates.
The activity of this surface has been extensively studied both experimentally and theoretically  with focus on surface  point defects such as bridging oxygen vacancies ($V_O$), hydroxyl groups (OH$_{br}$) or Ti interstitials in the near surface region.
However, despite their abundance at the rutile (110) surface \cite{Diebold1998137,new_steps,Lira2011}, systematic theoretical studies of the structure and the activity of extended defects such as step edges have never been reported before.

In this work, we report that reconstructed and reduced \oneone\ step edges on rutile \Tio(110) are mild trapping sites for the dissociation of small molecules such as water and methanol.
Starting from the bulk truncated structures of the two most stable  \oneone\ and \zero\ step edges, the structures of which are shown in Fig.~\ref{fig:surface}, we apply a systematic search for any other possible structure employing an automated genetic algorithm (GA) sampling up to 200 different candidates for each step edge.
Indeed, for the \oneone\ step we find a new configuration that is more stable with respect to the bulk truncated one.
When the new structure is considered, we find that oxygen vacancies along step edges have a low formation energy, meaning that this step is easily reducible.
We further demonstrate how water and methanol dissociate on reduced and reconstructed \oneone\ steps with binding energies of about 0.5~eV lower with respect to the highly reactive bridging $V_O$.
Our findings are in agreement with published water temperature programmed desorption (TPD) experiments and we propose a new assignment for methanol TPD measurements.

\section{Methods}

\subsection{Computational details}

The calculations were performed at the Generalized Gradient Approximation (GGA) level of DFT using the Perdew-Burke-Ernzerhof  (PBE) exchange correlation functional \cite{PBE} as implemented in the Grid-based Projector Augmented-Wave (GPAW) program \cite{gpaw,gpaw_review}.
Electrons are described using the projector augmented wave (PAW) method in the frozen core approximation \cite{PAW}.

The \zero\ and \oneone\ steps have been modeled with vicinal rutile (230) and (451) surfaces, respectively.
Periodic boundary conditions have been applied only along the surface plane $(x,y)$, whereas a vacuum of at least 7~\AA\ between the surfaces
and the walls of the cell was set.
A grid spacing of no more than 0.20~\AA\ was used.
The area of the vicinal surface times the number of k-points in the Brillouin zone is always bigger than 640~\AA$^2$.
A four electron setup (4{\it e}) has been used for Ti atoms, which consider the 3{\it d} and 4{\it s} electrons in the valence shell.
The equilibrium lattice parameters of the bulk rutile phase are {\it a} = 4.71~\AA\ and {\it c} = 3.01~\AA.

The graph in Fig.~\ref{fig:cicle} has been calculated on a Ti$_{37}$O$_{74}$(451) slab.
We have checked the accuracy of our calculations with the 4$e^-$ setup by employing a 12 electron setup (12$e^-$),  which includes also the 3{\it s} and the 3{\it p} electrons in the valence shell.
The equilibrium lattice parameters calculated for the 12$e^-$ setup are {\it a} = 4.65~\AA\ and {\it c} = 2.96~\AA.
The results using the 4$e^-$ and 12$e^-$ setups shows good correspondence ({\it cf.} caption of Fig.~\ref{fig:cicle}).

Oxygen vacancies have been created starting from a stoichiometric Ti$_{118}$O$_{236}$(451) slab, which corresponds to a (2$\times$1) unit cell.
A (451)--(1$\times$1) unit cell has been used to simulate a high concentration of oxygen vacancies.
Such configuration results in a step $V_O$ every O row or one $V_O$ per $\sim$7~\AA.

Adsorption of water and methanol on a \Tio(110) terrace has been preformed on a four TiO$_2$ trilayer slab using a (4$\times$2) unit cell.
The slab is asymmetric with the adsorbate on one side only.
The formation energy of a bridging $V_O$ with respect to an O$_2$ molecule in the gas phase is 3.26~eV.
The adsorption of water and methanol on a step oxygen vacancy has been modeled in a Ti$_{59}$O$_{117}$(451) slab.

\subsection{Genetic Algorithm}

When investigating a possible step reconstruction the atoms in the two rows of TiO$_2$ closest to the step are considered the target atoms for an optimization with a newly implemented genetic algorithm (GA) \cite{Lasse2011}. 
The GA starts from a population of 15 randomly generated candidates.
A fixed population size of the 15 best candidates is maintained throughout. 
To preserve the stoichiometry in the pairing processes each atomic species is paired separately with the cut-and-splice operator \cite{Deaven1995}. 
Mutations are introduced to ensure a diverse population with the primary mutation being a permutation of half of the atomic numbers in 30\% of the newly generated candidates. 
A similarity criterion based both on energy differences and a pair distribution function is used to limit the number of duplicates in the population. 
It is however not possible to fully hinder duplicates and the GA search is considered converged when small variations of the best candidate start dominating the population.
In order to sample a large number of structures we have used a faster approach during the DFT optimization step in the GA.
This method uses a Linear Combination of Atomic Orbitals (LCAO) to describe the wave functions  \cite{lcao_gpaw}.

\section{Results and Discussion}

\subsection{Step Structures}

\begin{figure}[!tb]
\begin{center}
\includegraphics[%
  width=0.48\textwidth,
  keepaspectratio]{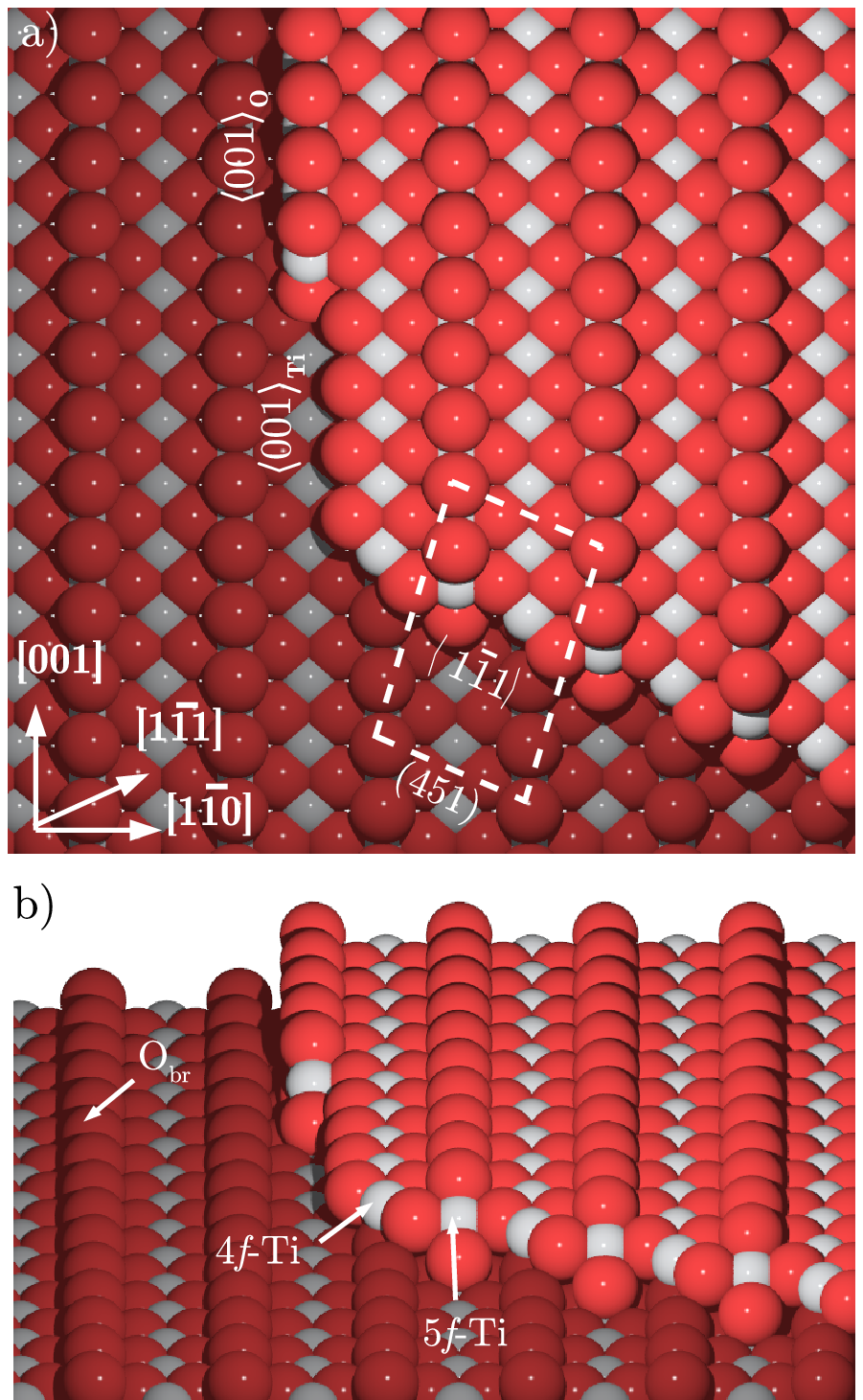}
\end{center}
\caption{\label{fig:surface}
(a) Top and (b) tilted side view of the \Tio(110) surface with bulk truncated \zero\ and \oneone\ step edges.
 }
\end{figure}

Figure \ref{fig:surface} illustrates the rutile \Tio(110) unreconstructed surface with bulk truncated steps running along the $[001]$ and $[1\bar11]$ directions.
Indeed, typical STM images show that terraces are bound by monoatomic \zero\ and \oneone\ steps of about 3.2~\AA\ height \cite{Diebold1998137}.
Steps that run along other directions are not observed under normal conditions  \cite{new_steps}.
In the present work, we have performed individual searches for the optimum structures of steps along the  $[1\bar11]$ and  $[001]$ directions. 
We first report the result for the \zero\ step for which we find the most stable step structure to be the bulk truncated one.

{\it \zero\ step edges}: Unreconstructed \zero\ step edges  can be created with two different terminations: O$_{br}$ rows and Ti troughs, \zeroa\ and  \zerob\ respectively, see Fig.~\ref{fig:surface}.
Based on STM images Diebold {\it et al.} \cite{Diebold1998137} have proposed that \zero\ steps are  O$_{br}$ terminated. 
We have applied to the \zero\ steps our GA search in order to find more stable structures than the bulk truncated ones.
However, after sampling up to 140 configurations, the two bulk truncated structures are recovered as the most stable ones.
From our calculations the \zeroa\ step is more stable by less than 0.05~eV per unit cell.
However, this energy difference is at the limit of our precision and therefore we are not able to confirm the experimental observation of \zeroa\ step to be the most stable one.

{\it \oneone\ step edges}: We now move to \oneone\ step edges that are expected to be present in a higher concentration \cite{Diebold1998137}. 
In Fig.~\ref{fig:surface}a the (451) surface unit cell used in the search for the most stable \oneone\ step is shown. 
The cell offers the smallest possible periodicity along the step direction, and separates the steps by about 13~\AA\ via flat \Tio(110) terraces containing four \Tio(110) unit cells in each row.  
A view at the (451) unit cell  in Fig.~\ref{fig:surface}a reveals that the step edge contains two inequivalent Ti atoms, five- and fourfold (5{\it f}- and 4{\it f}-) coordinated (Fig.~\ref{fig:surface}b), meaning that in order to add one row to the step and recover the same step structure two \Tio\ units must be added. 
This in turn means that the most stable step may either result from steps having an odd or an even number of such added \Tio\ units. 

\begin{figure}[!tb]
\begin{center}
\includegraphics[%
  width=0.47\textwidth,
  keepaspectratio]{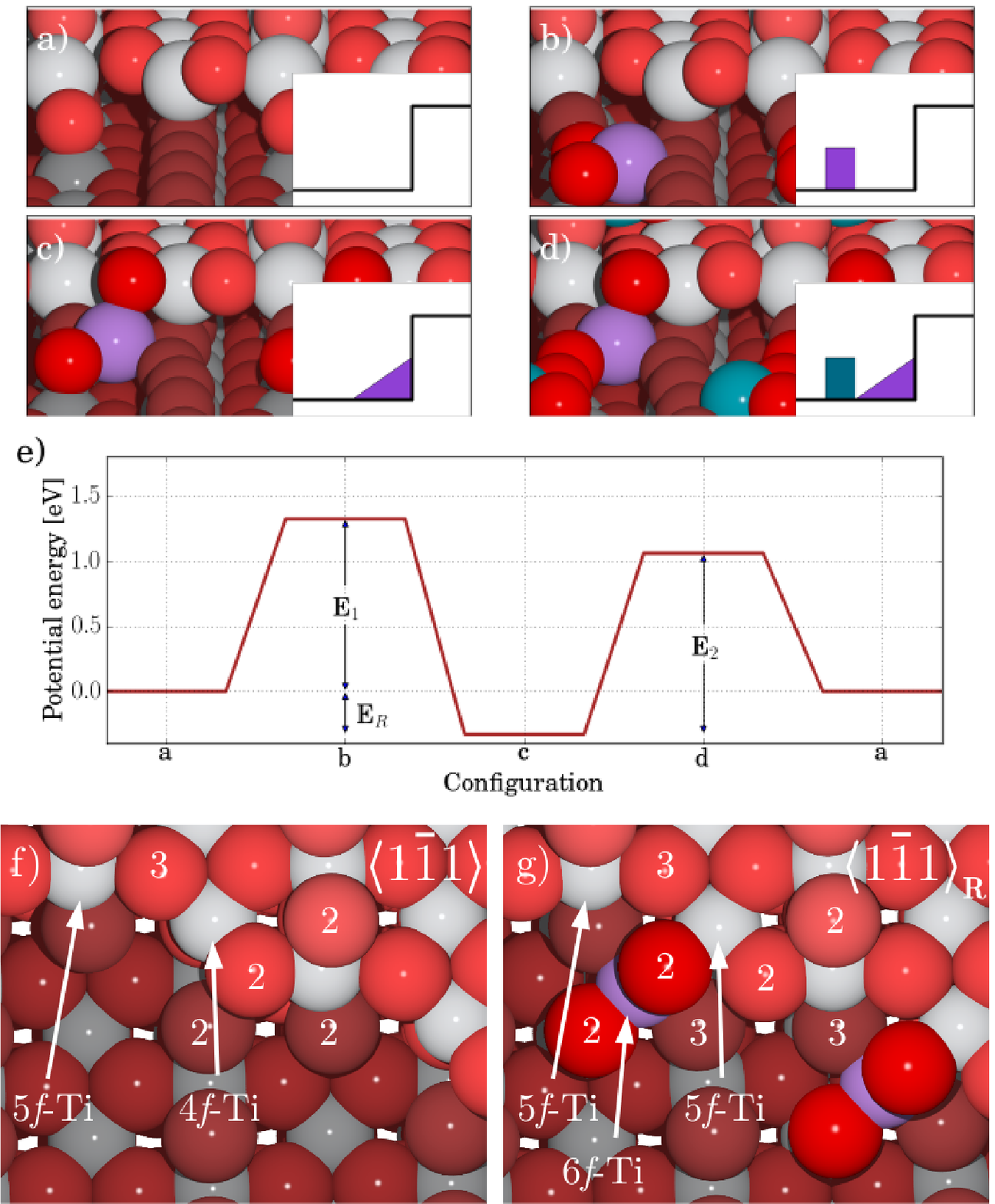}
\caption{\label{fig:cicle}
Growth of a \oneone\ step edge when going from (a) a bulk truncated \oneone\ step to the same structure but with two extra \Tio\ units.
Configuration (c) represents the most stable structure, \oneoner.
(e) Potential energy diagram for the six structures. 
The energy values calculated with a 12$e^-$ setup are: E$_1$=1.49~eV; E$_R$=$-$0.19~eV; E$_2$=1.40~eV.
(f-g) Top view of the \oneone\ and \oneoner\ step edges. 
The coordination numbers of the Ti and the O atoms are indicated counting Ti--O bonds of length up to 2.5~\AA.
}
\end{center}
\end{figure}

\begin{figure*}[!tb]
\begin{center}
\includegraphics[%
  width=0.98\textwidth,
  keepaspectratio]{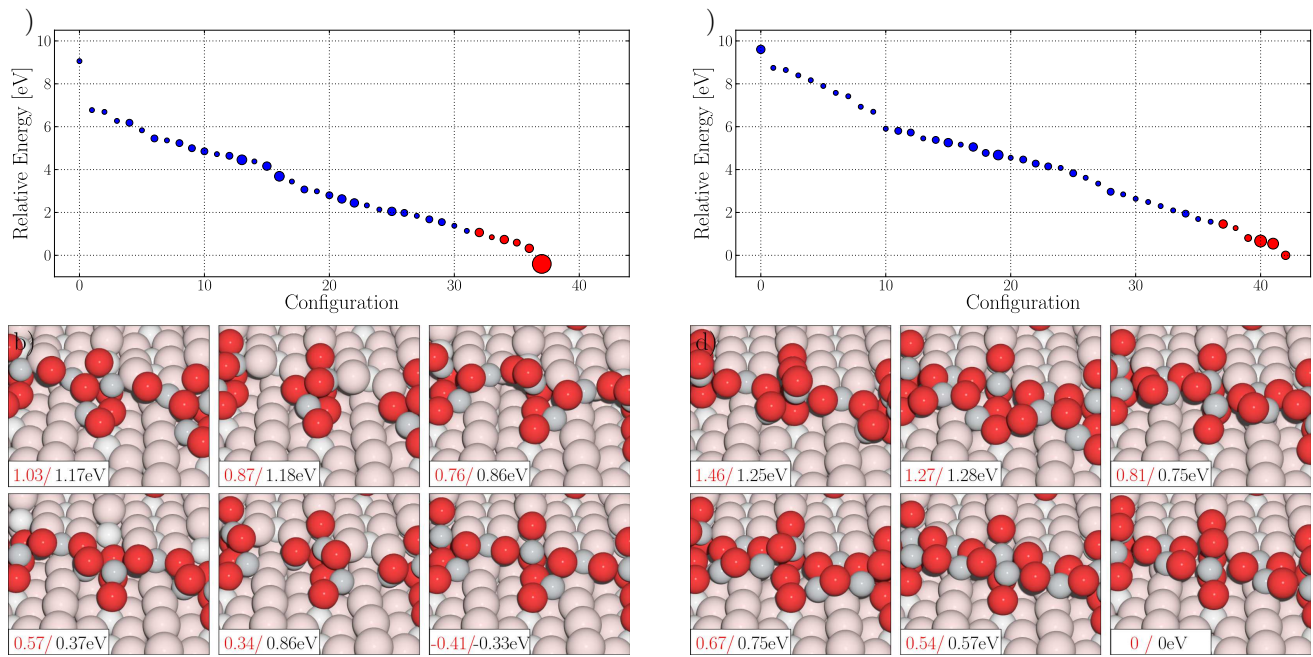}
\caption{\label{fig:GA}
Energy distribution for all the configurations sampled by the GA for the (a) \oneoner\ and (c) \oneone\ step edges.
The zero in energy corresponds to the bulk truncated \oneone\ structure.
The size of the dots is proportional to the number of times the configuration has been successfully discovered.
(b,d) Ball models of the best six structures for the \oneoner\ and the \oneone\ step edges.
The TiO$_2$ units included in the GA are in highlighted colors.
Energy values calculated with the LCAO (red) and the grid-based (black) methods are reported.
}
\end{center}
\end{figure*}

The growth of the \oneone\ step is illustrated in Fig.~\ref{fig:cicle}.
Starting from the bulk truncated step edge, Fig.~\ref{fig:cicle}a, one unit of \Tio\ is adsorbed on the terrace in the structure reported by Park  {\it et al.} \cite{PhysRevB.75.245415} and Wendt  {\it et al.} \cite{Wendt27062008}, Fig.~\ref{fig:cicle}b. 
With respect to bulk TiO$_2$ hosting the TiO$_2$ on the terrace is unfavorable by about 1.3~eV (E$_1$=1.32~eV). 
Moving the \Tio\ unit to be part of the \oneone\ step and after the search with the GA involving 194 structural candidates, the structure of Fig.~\ref{fig:cicle}c resulted as the most stable. 
Here, the TiO$_2$ unit is incorporated into the step edge and 1.65~eV is gained rendering this configuration the thermodynamically preferred state of the \oneone\ step edge under condition of stoichiometry by 0.33~eV (E$_R$).
We shall refer to this structure as \oneoner\ and return to the question of non-stoichiometric conditions in Section~\ref{sec:reduced}. 
One characteristic of the \oneoner\ structure is that the Ti atom added to the step edge (purple in Fig.~\ref{fig:cicle}c) appears at a height, 2.16~\AA, which is intermediate between that of the two terraces forming the edge, $\equiv 0$~\AA\ and 3.26~\AA.
In Fig.~\ref{fig:GA}a we report the energy of all the structures sampled by the GA.
The six most stable ones are illustrated in Fig.~\ref{fig:GA}b.

To complete the growth cycle and prove the consistency of our approach, a 2$^\mathrm{nd}$ TiO$_2$ unit was added on the terrace, Fig.~\ref{fig:cicle}d, which costs about 1.4~eV (E$_2$=1.39~eV). 
Moving this TiO$_2$ species (cyan color in Fig.~\ref{fig:cicle}) to the step edge and employing our search with the GA covering more than 100 structural candidates, the bulk truncated edge structure, Fig.~\ref{fig:cicle}a, was recovered.
The energy of the structures sampled by our GA within 10~eV are reported in Fig.~\ref{fig:GA}c, and the most stable structures are shown in Fig.~\ref{fig:GA}d.

We emphasize that the two structures, Fig.~\ref{fig:cicle}a and \ref{fig:cicle}c, have resulted directly from the GA search without any constraints or initial guesses of the configurations ({\it cf.} Fig.~\ref{fig:GA}), confirming the high stability of such step edges and proving the reliability of our approach.

We now address the reasons for the particular high stability of the \oneoner\ step with respect to the \oneone\ step.
The two structures for the steps running along the  $[1\bar11]$ direction are shown enlarged in Fig~\ref{fig:cicle}f-g.
From the structure reported in Fig.~\ref{fig:cicle}g we note that the two oxygen atoms of the extra \Tio\ unit, are in their bulk positions and are twofold coordinated.
The extra Ti atom is sixfold coordinated in an octahedral site at half height between the upper and the lower terrace.
Furthermore, the new in-plane O atom on the upper terrace establishes the terrace-like fivefold coordination of the 4{\it f}-Ti atom along the bulk truncated \oneone\ step.
The new \oneoner\ step that we have discovered thus optimizes the coordination of all Ti atoms resulting in a very stable structure.

\subsection{Reduced Step Edges}\label{sec:reduced}

Having establish a new structure for the \oneone\ steps we now turn to the study of oxygen vacancies ($V_O$) formed along  such step edges.
A reduced \Tio\ surface typically contains bridging $V_O$ in a concentration from 2 to 10\%~ML which are easily imaged by STM as bright protrusions along the dark O$_{br}$ rows.
The presence of $V_O$ at step edges with different coordination with respect to bridging $V_O$ is intriguing and a different activity can be expected.
Indeed, we find that creating a $V_O$ along this  step is easier compared to the creation of a $V_O$ on the terrace.

\begin{figure}[!tb]
\begin{center}
\includegraphics[%
  width=0.48\textwidth,
  keepaspectratio]{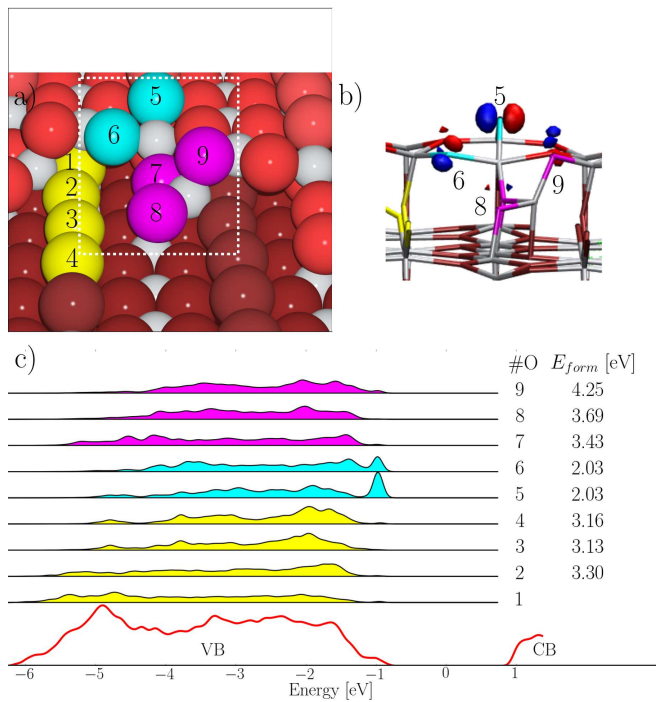}
\caption{\label{fig:vac}
(a) \oneoner\ step edge. 
The O atoms where $V_O$ are created are marked in yellow, cyan and purple.
(b) Highest occupied Kohn-Sham orbital. The isodensity value displayed is $\pm$0.4~\AA$^{-3/2}$.
(c) Total density of states and 2{\it p} projected density of states of the 9 O atoms marked in (a).
On the right side of the plot the corresponding $V_O$ formation energies are reported.
}
\end{center}
\end{figure}

In Fig.~\ref{fig:vac} the formation energies for a series of $V_O$ created by removing different O atoms in our model of the \oneoner\ step are reported.
Oxygen atoms numbered 2, 3 or 4 are representative of terrace O$_{br}$ atoms.
A $V_O$ created by removing one of these atoms costs about 3.1-3.3~eV.
The oxygen atoms that are marked from 5 to 9 are instead characteristic of the \oneoner\ step.
The results show how steps are clearly highly reducible with $V_O$ at position 5 having the lowest formation energy, 2.03~eV.
Also removing the O atom number 6 costs the same amount of energy but in this case the structure undergoes a strong relaxation and a $V_O$ in position 5 is created.
We explain the low formation energy of a $V_O$ in position 5 in terms of electronic states associated with the O atoms prior to the $V_O$ formation.
In Fig.~\ref{fig:vac}c the projected density of states of the 2{\it p} orbitals of the O atoms of interest are shown.
It is striking that the two O atoms that are easy to remove are the ones that contribute to the top of the valence band with a well defined state of  2{\it p} character,  Fig.~\ref{fig:vac}c.
This state is localized on atoms number 5 and 6 as depicted in Fig.~\ref{fig:vac}b.
Moreover, it has an anti-bonding character with a nodal plane between the two atoms, explaining the lower energy cost of removing any of them.
The quoted $V_O$ formation energies apply to the removal of every 2$^\mathrm{nd}$ atom of the given type along the step edge. 
When increasing the coverage and removing one O atom in every O$_{br}$ row  the $V_O$ formation energies remain low (2.25~eV for site no.~5) indicating that \oneoner\ steps can accommodate a high density of oxygen vacancies, one $V_O$ per $\sim$7~\AA.

Based on these results, we conclude that a reduced \Tio(110) sample that shows even a low concentration of bridging $V_O$ at the terraces, must accommodate a high number of $V_O$ along \oneoner\ step edges.

\subsection{Adsorption of Water and Methanol}

\begin{figure}[!tb]
\begin{center}
\includegraphics[%
  width=0.48\textwidth,
  keepaspectratio]{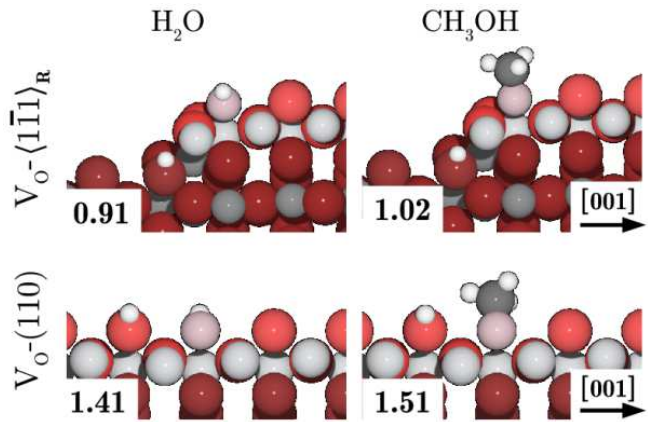}
\caption{\label{fig:water}
Dissociative adsorption for a water and a methanol molecule in an oxygen vacancy at step (top) and in a bridging oxygen vacancy (bottom).
}
\end{center}
\end{figure}

Given the structures for reduced steps running along the $[1\bar11]$ direction we have tested the activity of such extended defects towards the adsorption of water and methanol and compared to the activity of the well known reduced and flat (110) terrace, Fig.~\ref{fig:water}.
Both water and methanol molecules dissociate and bind strongly at bridging $V_O$  (Fig.~\ref{fig:water}, lower part) in agreement with previous theoretical and experimental works \cite{Brookes2001,Wendt2006,Armas2007,Henderson1999}.
In contrast to a molecular adsorption along Ti troughs characterized by lower binding energies, bridging $V_O$ are considered the only trapping site where these molecules can dissociate.
In Fig.~\ref{fig:water} the most stable configurations for the adsorption of H$_2$O and CH$_3$OH into a step $V_O$ are illustrated.
Both molecules dissociate with binding energies of 0.91 and 1.02~eV for water and methanol, respectively.

The results for dissociation of water at step edges are in very good agreement with earlier water TPD experiments that show a recombinative desorption peak at temperatures lower than the one at vacancy sites \cite{Henderson1998,Henderson2003}.
However, the presence of this peak in H$_2$O TPD seems to be dependent on the experimental condition (e.g. see TPD spectra in Ref.~\cite{Hugenschmidt1994,Henderson1998} compared to the one in Ref.~\cite{Wu2008} where this intermediate peak is not discussed).
In addition,  hydroxyl groups at step edges resulted from water dissociation have been reported to be observed by Suzuki {\it et al.} \cite{Suzuki2001}.

Our results suggest a similar behavior of water and methanol.
We predict methanol to dissociate at reduced \oneoner\ steps.
This result is compatible with methanol TPD experiments performed by Henderson {\it et al.} \cite{Henderson1998a}.
The TPD spectra for methanol adsorbed on a \Tio(110) surface shows three distinct states at 295, 350 and 480~K.
The low and the high temperature peaks are assigned to molecular methanol adsorbed along Ti troughs (E$_b$=0.84~eV) and to dissociative methanol adsorbed at bridging $V_O$ sites  (E$_b$=1.51~eV, Fig.~\ref{fig:water}), respectively.
We propose the methanol dissociated at vacancy sites along step edges, the configuration shown in Fig.~\ref{fig:water}, to be responsible of the third peak at 350~K (E$_b$=1.02~eV).
This assignment is compatible with the appearance of this peak at low coverage and with the authors' conclusion that methanol desorbing  at this temperature is formed from recombination of dissociated methoxy species.

\section{Conclusions}

In summary, we have provided a clear understanding of the local structure of step edges on rutile \Tio(110) surface.
An automated genetic algorithm was employed leading to the finding of a new structure for the \oneone\ step edge that differed from the bulk truncated one. 
Knowing the correct structure of step edges is of fundamental importance and can have many consequences for several reasons.
Here, we have shown that oxygen vacancies can be extremely stable at step edges.
A reduced rutile sample thus hosts not only point defects, like oxygen vacancies or Ti interstitial, but also highly reduced step edges.
This conclusion can easily be extended to other reducible metal oxide system and can strongly influence the activity of the near step region.
For example, water and methanol molecules dissociate at reduced \oneone\ step edges as in bridging $V_O$, but with lower binding energies.
This implies that while point defects at terraces become polluted, point defects at step may engage in a catalytic cycle.
The results for water are in good agreement with earlier experimental results where it has been speculated that step edges may play a role in the water dissociation \cite{Suzuki2001,Henderson1998,Henderson2003}.
For methanol adsorption we reinterpret previous experimental results  \cite{Henderson1998a,Henderson1999} and propose a similar behavior for higher alcohols.

Our findings also suggest the possibility of tuning the activity of the rutile \Tio(110) surface allowing to decrease the temperature needed for the desorption of strongly bounded species.
For example, changing the experimental conditions or the surrounding environment can result in a different morphology of the surface \cite{Jak2002,Aschauer2011} and in particular in a different ratio between flat (110) terraces and step edges.
Moreover, vicinal surfaces that expose a high concentration of step edges oriented along specific directions can be employed \cite{Tegenkamp2009}.

\section{Acknowledgment}

This work has been supported by the Danish Research Councils and the Danish Center for Scientific Computing.

\end{document}